# Multi-Organ Cancer Classification and Survival Analysis

**Stefan Bauer, Nicolas Carion, Joachim M. Buhmann**
Department of Computer Science
ETH Zurich, Switzerland
`{bauers, ncarion} @inf.ethz.ch`
`jbuhmann @inf.ethz.ch`

**Peter Schüffler, Thomas Fuchs**
Memorial Sloan Kettering Cancer Center
New York, USA
`{schueffp, fuchst}@mskcc.org`

**Peter Wild**
Institute of Surgical Pathology
University Hospital Zurich
`peter.widl@usz.ch`

## Abstract

Accurate and robust cell nuclei classification is the cornerstone for a wider range of tasks in digital and Computational Pathology. However, most machine learning systems require extensive labeling from expert pathologists for each individual problem at hand, with no or limited abilities for knowledge transfer between datasets and organ sites. In this paper we implement and evaluate a variety of deep neural network models and model ensembles for nuclei classification in renal cell cancer (RCC) and prostate cancer (PCa). We propose a convolutional neural network system based on residual learning which significantly improves over the state-of-the-art in cell nuclei classification. Finally, we show that the combination of tissue types during training increases not only classification accuracy but also overall survival analysis.

## 1 Introduction

To facilitate automated cancer diagnosis and prognosis, computational pathology is providing fully automated image analysis pipelines, e.g. [5], [4] or [13]. While these results already match or surpass the classification accuracy of expert pathologists [1], they require extensive feature engineering and extensive expert labels for specific cancer types. The ongoing success in machine learning and computer vision demonstrates the remarkable learning abilities of deep networks for image recognition [e.g. 12]. Deep learning algorithms have already been successfully applied in computational pathology, e.g. for the *ICPR 2012 Contest on Mitosis Detection in Breast Cancer Histological Images* [2] and similarly for the *MICCAI 2013 Grand Challenge* [17].

Our main motivation is to investigate the performance of networks trained from scratch with fixed parameters to see the transfer of learned concepts from one organ to the next. To the best of our knowledge, there exists only very limited information on common features of cancer cells from different organs, thus potentially requiring the tailoring of current frameworks to specific cancer scoring tasks.

In 2 we describe both datasets and provide benchmark comparisons in 3. Given the typical small data sets in computational pathology, we focus on data augmentation in 3.1 and study the learning of neural networks of different depths and ensembles when trained from scratch in 3.2 and 3.3, as well as the transfer of trained networks from one organ to the next (with fixed parameters) and on the joint data set with two, three or four classes in 3.4. We additionally validate our results by showing that not only the classification accuracy is improved but also overall survival analysis in 4. Our approaches significantly improve over state-of-the-art in cell nuclei classification and suggest common patterns in renal cell carcinoma (RCC) and prostate cancer (PCa). In addition we provide our code and data for bench-marking and future research in computational pathology.

## 2 Data

**Renal Cell Carcinoma (RCC)** belongs to the 10 most common cancers in western societies' mortality [7]. Clear cell renal cell carcinoma (ccRCC) is a common subtype of RCC occurring on cells with clear cytoplasm. Since this cancer develops metastases in a very early stage (commonly before diagnosis), the prognosis for RCC patients is usually pessimistic [16]. Tissue microarrays (TMA) serve as an important tool for molecular biomarker discovery, since they enable the screening of dozens or even hundreds of specimen simultaneously. Our data basis are eight ccRCC TMA images. Each image is fully labeled by two pathologists, indicating location and class of all malignant and benign cell nuclei. From 1633 found nuclei, the two pathologists agreed with the labels on 1272. These 1272 well-labeled nuclei (890 benign and 382 malignant) were extracted as patches of size 78x78 pixels centered at labeled nuclei and serve as our original study data (see Fig 1). This dataset was first published and analysed in [5] and serves jointly with [13] as a comparison.

**Prostate cancer (PCa)** is one of the most common cancer types in western male society. It is the second most frequently diagnosed cancer for human males worldwide, and the sixth leading cause of cancer related death [9]. However, research is ongoing for the development of specific biomarkers for the early diagnosis and the deeper understanding of PCa [6]. We incorporate six new TMA images of PCa patients, twice labeled by two pathologists. From 1195 detected nuclei, they agreed on the label of 826 (207 benign, 619 malignant).

## 3 Experiments

We used the Caffe library [10] to train variants of small Cifar10 [11], AlexNet [12], ImageNet [3] and googlenet [15] like deep networks. Given our small data set we additionally implemented the newly developed residual networks [8], which outperforms previous approaches in the ImageNet competition. A residual network with 18 layers is denoted by ResNet18 and one with 34 layers by ResNet34. All larger models e.g. ImageNet quickly led to overfitting and poor results due to the small sample size. Additionally, we tested the inclusion of custom weights in the cost function e.g. for AlexNet [12], in order to overcome class biases. However, experiments showed that this compensation does not improve the error rates. While the code for the best performing ResNet's is already

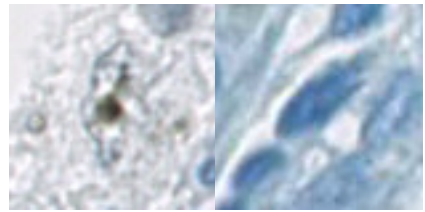

**(a)** Renal cell  **(b)** Prostate cell

**Figure 1:** Examples of $78 \times 78$ patches.

provided by Facebook https://github.com/facebook/fb.resnet.torch, we provide our code for the customized data augmentation and both data sets for bench-marking and future research. The best classifier using hand-crafted feature engineering, achieves a classification accuracy of $83\%$ for RCC [13], which is as good as the manual annotation: the inter-pathologist accuracy for classification of 1633 renal clear cell carcinoma nuclei is $80\%$. Replicating the approaches in [13] for PCa, these values even increase up to $90\%$. The automated staining estimation pipeline is implemented in the free Java program TMARKER [14]. The reported performance measures are *recall*, *precision*, *F1-score* and *support*.

### 3.1 Data augmentation

We randomly split the data into $80\%$, $10\%$ and $10\%$ for training, testing and validation. Due to the computational cost we only apply a one fold cross validation and split and average the data only once. For comparison for some nets the results of a double split experiment are reported. Given the low number of samples available one focus of our work is data augmentation and we apply the following techniques to the training set while averaging over all predictions for the validation set: firstly, the nucleus patch is *scaled down* to a randomly chosen size in $[64 : 78]$. After that, we select uniformly at random a *cropping* of size $64 \times 64$, and we *mirror* it with probability $1/2$. Since only the shape, and not the orientation or the color of the nuclei is discriminative for classifying it as malignant or benign [13], we also apply a *rotation* by a random angle between 0° and 360°, and in addition *grayscaling*. Each picture is randomly perturbed 50 times, giving alltogether 60T pictures in the RCC and 40T for the PCa dataset.

### 3.2 RCC

Using a random partition with 80% for training and 10% for testing and validation, the performance is comparable or significantly better to the hand-crafted approach in [13] with a score of 83% (2b). While the overall performance of the Cifar10 net is comparable to the hand-crafted approach in [13], it apparently has difficulties with the prediction for the malignant cells as indicated by low precision and recall. Due to the limited number of samples in the validation set a multiple random partitions into training, testing and validation might already reduce the chance of an unfavorable validation set, as shown for the residual networks.

| Data | *Precision* | *Recall* | *F1* | *Support* |
|---|---|---|---|---|
| malignant | 0.79 | 0.86 | 0.83 | 44 |
| benign | 0.93 | 0.88 | 0.90 | 84 |
| Avg./Tot. | 0.88 | 0.88 | 0.88 | 128 |

(a) ResNet18 on RCC

| Data | *Precision* | *Recall* | *F1* | *Support* |
|---|---|---|---|---|
| malignant | 0.79 | 0.68 | 0.73 | 44 |
| benign | 0.84 | 0.90 | 0.87 | 84 |
| Avg./Tot. | 0.83 | 0.83 | 0.82 | 128 |

(b) ResNet34 on RCC

**Figure 2:** Renal cell carcinoma (RCC) performance comparison of residual networks (ResNets) with different number of layers.

### 3.3 PCa

In addition to the renal cell carcinoma data, we tested the different deep learning architectures on *MIB-1* stained prostate cancer TMAs. The performance for both the residual networks with 18 and 34 layers is close to the intersection of two pathologists Fig. 3a and Fig. 3b. However both nets misclassify a different patch. An ensemble of both nets with equal weights only misclassifies one of the two patches, since the confidence of the residual net with 34 layers is high enough to overcome the wrong label of the residual net with 18 layers. When trained on the combined data of RCC and PCa, all pictures (and thus the two pictures as well) are correctly classified by both ResNet18 and ResNet34 (see Section 3.4).

| Data | *Precision* | *Recall* | *F1* | *Support* |
|---|---|---|---|---|
| malignant | 0.99 | 1.00 | 0.99 | 69 |
| benign | 1.00 | 0.93 | 0.96 | 14 |
| Avg./Tot. | 0.99 | 0.99 | 0.99 | 83 |

(a) ResNet18 on PCa

| Data | *Precision* | *Recall* | *F1* | *Support* |
|---|---|---|---|---|
| malignant | 0.99 | 1.00 | 0.99 | 69 |
| benign | 1.00 | 0.93 | 0.96 | 14 |
| Avg./Tot. | 0.99 | 0.99 | 0.99 | 83 |

(b) ResNet34 on PCa

**Figure 3:** Performance of networks with different layers for prostate cancer. While only one sample is misclassified for both networks, it is a different patch each time.

### 3.4 Multi-organ RCC and PCa data

For the multi-organ cancer classification we conducted two different kinds of experiments: first, we run a four class classification with two classes per organ (i.e. malignant and benign for prostate and malignant and benign for renal cells); second, we only used two classes (i.e. malignant and benign). While the evaluation on the RCC set decreases the accuracy on the validation set to 80% (see 4b), the true performance might be higher. The training accuracy is around 86% and the survival analysis in Section 4 shows a significant improvement for the RCC data. Likewise, the performance on the PCa data is improved since now no sample is miss-classified and we exactly replicate the results from the intra-pathologist agreement. We regard it as a positive feature that: *No cell of one organ was labeled as cell of a different organ*. Similarly, the two-class residual networks with 18 and 34 layers trained on the combined data set have 100% accuracy on the PCa data while the performance for the RCC drops to 80%. In addition to the residual networks, the Cifar10 model trained on the joint set and validated on the PCa data achieves very good results.

## 4 Survival Analysis

In addition to classification, we tested our approaches on follow-up survival data on 132 RCC patient. In RCC, the staining estimate of the proliferation protein *MIB-1* is corelated with the overall survival

| Data | *Precision* | *Recall* | *F1* | *Support* |
|---|---|---|---|---|
| malignant | 0.67 | 0.84 | 0.75 | 44 |
| benign | 0.90 | 0.79 | 0.84 | 84 |
| Avg./Tot. | 0.82 | 0.80 | 0.81 | 128 |

**(a)** ResNet18 trained on RCC and PCa and evaluated on RCC (benign and malignant)

| Data | *Precision* | *Recall* | *F1* | *Support* |
|---|---|---|---|---|
| malignant | 0.73 | 0.68 | 0.71 | 44 |
| benign | 0.84 | 0.87 | 0.85 | 84 |
| Avg./Tot. | 0.80 | 0.80 | 0.80 | 128 |

**(b)** ResNet18 trained on RCC and PCa and evaluated on RCC (four classes)

**Figure 4:** Performance for ResNet with 18 and 34 layers for a two class classification.

outcome. The staining estimate is the relative amount of stained cells *among the cancer cells* in the image. On 132 TMA images of RCC patients 130,899 cell nuclei have been detected earlier, as well as labeled as stained or not. For accurate staining estimation, we extract the nuclei as 78x78px patches and classify them into malignant or benign with the proposed models and compare the staining estimate to a trained pathologist. The patients have been stratified into two equally sized groups and the Kaplan-Meier survival estimator is plotted. The log rank test was used to test for significant survival differences between the patient groups (see Fig. 5). While the benefit for training on the combined data set of both organs was not evident by the performance measures in 3.4, the residual net with 18 layers outperforms the human pathologist (Fig. 5) and previous approaches [5] as indicated by a p-value of $0.006$ for the ResNet18 compared to a p-value of $0.038$ for the pathologist. While training on the combined set led to a significant gain for the residual network with 18 layers it does not help improving the model with 34 layers. This indicates that the smaller model finds a good balance between complexity and available data. Jointly with the insight that a four class classification does not lead to improved results compared to a two-class classification, we find evidence supporting common features for multi-organ cancer detection.

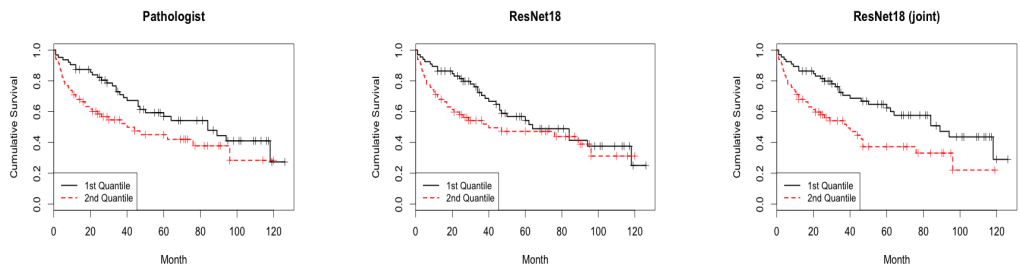

**Figure 5:** Kaplan-Meier estimators based on the manual estimates of a pathologist (**left**), the predictions of ResNet18 trained on RCC alone (**middle**), and the prediction of ResNet18 trained on the joint data set of RCC and PCa (**right**).

## 5 Conclusion

Histologic nuclei classification is a crucial precursor for a plethora of research tasks in computational pathology. In this paper we proposed a deep learning framework for application on *MIB-1* stained renal cell cancer and prostate cancer tissue microarrays. Our contributions are (i) developing and providing extensive data augmentation procedures (ii) the detailed evaluation of various state-of-the art convolutional neural network (CNN) architectures, (iii) the implementation of multi-organ prediction models, (iv) the evaluation of CNN ensemble models and (v) the application to survival analysis of RCC patients.

We are convinced that the proposed pipeline, together with the published code, image datasets, and survival information will serve as an useful and extensive benchmark for future computational research to the whole community.

**Acknowledgments**

This research was partially supported by the Max Planck ETH Center for Learning Systems and the SystemsX.ch project SignalX.